\documentclass{aa}

\usepackage{graphicx}
\usepackage{txfonts}
\usepackage{url}
\usepackage[colorlinks=true,citecolor=red]{hyperref}
\usepackage{amsmath}
\usepackage{color}
\usepackage{xcolor}
\usepackage{txfonts}
\usepackage{rotating}
\usepackage{float}
\usepackage{lscape}
\usepackage{longtable}
\usepackage{arydshln}
\usepackage{float}

\newcommand{\Msun}{M$_\odot$}

\newcommand{\Zsun}{Z$_\odot$}
\newcommand{\kms}{\,km\,s$^{-1}$} 
\newcommand{\orcid}[1]{\href{https://orcid.org/#1}{\includegraphics[width=10pt]{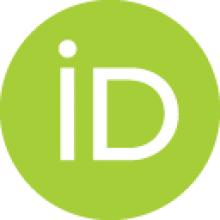}}}

\definecolor{yaleblue}{rgb}{0.2,0.5,0.95}
\definecolor{lava}{rgb}{0.81, 0.06, 0.13}
\definecolor{forestgreen}{rgb}{0.0, 0.45, 0.13}
\hypersetup{colorlinks=true, linkcolor=lava, urlcolor=forestgreen, citecolor=yaleblue}

\authorrunning{Guti\'errez et al.}
\titlerunning{CCSN ratios: observations vs predictions}

\begin{document}

\title{Relative frequencies of core-collapse supernovae as a function of metallicity: observations vs theoretical predictions}


\author{
Claudia P. Guti\'errez,\inst{1,2}\orcid{0000-0003-2375-2064}
\and
Llu\'is Galbany,\inst{2,1}\orcid{0000-0002-1296-6887}
\and
Joseph P. Anderson\inst{3}\orcid{0000-0003-0227-3451}
\and
Dimitris Souropanis\inst{4}\orcid{0000-0002-6548-5489}  
\and
Emmanouil Zapartas\inst{5,4}\orcid{0000-0002-7464-498X} 
\and
Luc Dessart\inst{6}
\and 
Rubina Kotak\inst{7}\orcid{0000-0001-5455-3653}
}
\institute{
Institut d'Estudis Espacials de Catalunya (IEEC), Edifici RDIT, Campus UPC, 08860 Castelldefels (Barcelona), Spain\\
\email{cgutierrez@ice.csic.es} 
\and
Institute of Space Sciences (ICE, CSIC), Campus UAB, Carrer de Can Magrans, s/n, E-08193 Barcelona, Spain
\and
European Southern Observatory, Alonso de C\'ordova 3107, Vitacura, Casilla 19001, Santiago, Chile
\and
Institute of Astrophysics, Foundation for Research and Technology–Hellas, GR-71110 Heraklion, Greece.
\and
Physics Department, National and Kapodistrian University of Athens, 15784 Athens, Greece.
\and
Institut d’Astrophysique de Paris, CNRS-Sorbonne Université, 98 bis boulevard Arago, F-75014 Paris, France
\and
Department of Physics and Astronomy, FI-20014 University of Turku, Finland
}


 
\abstract
{
Understanding supernova (SN) progenitors remains a major challenge in astrophysics, as it involves untangling the complex interplay between stellar physics (e.g., evolution,  binarity, explosion) and environments (e.g., metallicity, star formation rate). To address this, we present relative frequencies of core-collapse SNe (CCSNe) as a function of metallicity using two complementary samples: (i) all literature SNe that have associated host galaxy parameters (absolute magnitudes, stellar masses, and/or oxygen abundances); and (ii) SNe classified between 2019 and 2024 with host magnitude information, including distance-limited subsamples within 50 Mpc and 100 Mpc. 
We found that CCSNe from the literature sample are associated with luminous galaxies, reflecting both the higher stellar content of such systems and selection biases inherent to targeted surveys. In contrast, the distance-limited subsamples provide a less biased view, showing that hydrogen-rich SNe (SNe~II) are more commonly found in lower-luminosity galaxies than stripped-envelope SNe (SESNe).
Comparisons between the literature sample and distance-limited subsamples indicate that trends derived from global measurements remain consistent. For the SESNe-to-SNe~II ratios, we confirm a slight increase with metallicity, reflecting a higher fraction of SESNe in metal-rich environments. 
Comparison with theoretical predictions shows that models including either binary interactions or rotation can broadly reproduce the observed trends, although degeneracies remain, and no single scenario uniquely explains the data. Overall, our results provide observational constraints on massive-star evolution and highlight the key role of metallicity and binarity in shaping the observed diversity of CCSNe.
}

\keywords{supernovae: general; stars: massive, stars: binaries, stars: evolution
}

\maketitle
%

\section{Introduction}

Core-collapse supernovae (SNe) mark the deaths of massive stars (M$_{ZAMS}>8-10$ \Msun), whose cores can no longer support themselves against gravity once nuclear fuel is exhausted, leading to the collapse of the iron core and a subsequent powerful explosion. Core-collapse SNe (CCSNe) represent one of the most diverse and complex classes of transient phenomena, exhibiting a broad range of spectroscopic and photometric behaviours. Historically, they were divided into hydrogen-rich (Type~II) and hydrogen-poor (Type~I) events \citep{Minkowski41}. Additional subclasses were later introduced based on the presence or absence of specific spectral features. Events showing silicon lines are classified as thermonuclear Type Ia SNe, which originate in binary systems containing at least one white dwarf; therefore, SNe~Ia are not part of the CCSN population.  Among Type~I CCSNe, those displaying helium lines are classified as Type~Ib, whereas those lacking helium are designated as Type~Ic SNe \citep[e.g.][]{Filippenko97}. Within the Type~Ic group, a subset with exceptionally broad spectral lines ($\sim15000 - 30000$ \kms) is classified as Type Ic broad line \citep[SNe~Ic-BL; e.g.][]{GalYam17, Modjaz19}. A transitional subtype linking the hydrogen-rich and hydrogen-poor classes was designated as Type~IIb SN (SNe~IIb), which evolves from hydrogen-dominated spectra at early phases to helium-dominated ones at later times \citep{Filippenko88}.

The spectral differences among CCSNe subtypes are generally interpreted as indicators of the degree of envelope stripping experienced by their progenitors. For SNe~II, the progenitors undergo relatively low mass loss, allowing them to retain most of their hydrogen envelopes until explosion. In contrast, Type~IIb progenitors experience partial stripping, preserving only a thin layer of hydrogen; Type Ib progenitors are largely stripped of hydrogen, while Type Ic progenitors have lost most or all of both their hydrogen and helium envelopes \citep{Dessart20}. Although the dominant mechanism responsible for this mass loss remains debated, it may result from binary interaction \citep[e.g.][]{Podsiadlowski92, Nomoto95, Eldridge08, Ercolino24, Dessart24}, stellar winds \citep[e.g.][]{Woosley93, Vink00, Heger03}, or a combination of the two. 

Parameters such as rotation and metallicity are also thought to play significant roles in shaping this mass stripping process \citep[e.g.][]{Meynet00, Hirschi07, Yoon10, Long22}. In particular, it is well established that line-driven mass loss in massive stars depends on metallicity \citep{Maeder92, Nugis00, Vink01}. Therefore, single stars in metal-poor environments are expected to lose less mass and thus produce a higher fraction of SNe~II relative to SNe~Ibc \citep[e.g.][]{Meynet05, Eldridge08, Georgy09}. However, stellar rotation can enhance mass loss even in low-metallicity stars \citep[e.g.][]{Meynet06, Hirschi07}, introducing additional complexity in predicting the relative fractions of stripped-envelope SNe (SESNe).

When binary evolution is considered, the predictions change substantially. At low metallicities, models predict an increased rate of SESNe \citep[e.g.][]{Yoon12} and a broader parameter space for binary SN~IIb \citep{Yoon17, Sravan19, Souropanis25}. However, the fraction of SNe~Ic is still expected to decline, with the SN~Ic$/$SN~Ib ratio dropping to $\sim0.1$ at $\sim0.02$ \Zsun\ \citep{Yoon10}. More recent work by \citet{Souropanis25} suggests that primarily binary interactions drive pre-SN mass loss and determine the resulting stripped-envelope SN subtypes across a wide range of metallicities. As a consequence, the overall SESN-to-SN II ratio remains roughly constant across metallicities, although both the SN~Ib$/$SN~II and SN~Ic$/$SN~II rates decrease toward lower metallicities.

Observational evidence, however, remains inconclusive. Several studies \cite[e.g.,][]{Prantzos03, Prieto08, Boissier09, Graur17} report that the SESN-to-SN~II ratio increases with metallicity. In contrast, other works \citep{Anderson15, Kuncarayakti18} find this ratio to be nearly constant across metallicities. This apparent discrepancy likely arises from differences in both methodology and sample selection.

From a methodological perspective, the definition of the SESN class strongly affects the results. For instance, SNe~IIb are sometimes included among SESNe, among SNe~II, or excluded entirely, which naturally alters the inferred ratios. In addition, misclassifications can further bias these ratios \citep[e.g.][]{Gonzalez-Banuelos25}. The approach to estimating metallicity also varies: many studies reporting a metallicity dependence derive metallicities from global galaxy luminosities using the luminosity–metallicity relation \citep[e.g.][]{Tremonti04}, while others use direct measurements at the explosion sites.
Sample selection introduces an additional source of bias. Most existing samples under-represent faint galaxies, resulting in limited metallicity coverage. \citet{Arcavi10} showed that when CCSNe from dwarf galaxies are included, an increase of SNe~IIb and SNe~Ic-BL is observed.

The above differences in methodology and sample completeness highlight the complex interplay between metallicity, binarity, and other stellar parameters, and underscore the need for larger, more homogeneous datasets to disentangle their respective roles. In this paper, we investigate the relative frequencies of CCSNe as a function of metallicity by comparing a heterogeneous sample from the literature with a more uniform and less biased sample of SNe classified between 2019 and 2024. The paper is organised as follows. A brief description of the sample is provided in Section~\ref{sec:sample}. Section~\ref{sec:analysis} presents the analysis and results. In Section~\ref{sec:discussion} we present the discussion, while our summary is given in Section~\ref{sec:conclusions}. Throughout this work, we assume a flat $\Lambda$CDM universe, with a Hubble constant of $H_0=70$\,km\,s$^{-1}$\,Mpc$^{-1}$, and $\Omega_\mathrm{m}=0.3$.

\section{SN sample and galaxy measurements}
\label{sec:sample}

\begin{table}
\small
\centering
\caption{Number of objects with galaxy measurements obtained from the literature.}
\label{tab:litsample}
\begin{tabular}{lccccc}
\hline
\hline
SN type   & M$_g^{host}$$^{\star}$ & M$_*$  &  N2  &   O3N2  \\
          &          [mag]          & \Msun\ & [dex] & [dex] \\
\hline
SNe~II    &       807           &   797    &  232  &  129  \\
SNe~IIb   &        95           &    85    &   23  &   18  \\
SNe~Ib    &        95           &    83    &   43  &   42  \\
SNe~Ic    &       110           &   116    &   78  &   70  \\
SNe~Ic-BL &        61           &    47    &   29  &   30  \\
SNe~Ibc   &         6           &    67    &    7  &    7  \\
\hline
\end{tabular}
\begin{list}{}{}
\item \textbf{Notes:} 
\item $^{\star}$ For the host absolute magnitude, we use either M$_g^{host}$ or M$_B^{host}$. When both measurements are available, we adopt M$_g^{host}$ as the reference value.
\item - In Figure~\ref{fig:histograms}, SESNe include SNe~IIb, Ib, Ic, Ic-BL, Ibc. 
\end{list}
\end{table}

The data used in this analysis comprises two samples: (i) SNe from the literature with available measurements of metallicity, host absolute magnitudes (M$_g^{host}$ or M$_B^{host}$) and/or stellar mass (M$_*$); and (ii) SNe classified between 2019 and 2024 and reported to the Transient Name Server (TNS\footnote{\url{https://www.wis-tns.org/}}). 

For the \textit{literature sample,} we compiled galaxy measurements from sample analyses including,  \citet{Modjaz08, Svensson10, Arcavi10, Leloudas11, Modjaz11, Sanders12, Taddia13, Stoll13, Perley13, Kuncarayakti13, Lunnan14, Bianco14, Taddia15, Kruhler15, Anderson15, Anderson16, Galbany16, Perley16a, Taddia16a, Taddia16b, Chen17, Schulze18, Kuncarayakti18, Galbany18, Gutierrez18, Taddia19, Modjaz20, Schulze21} and \citet{Grayling23}, resulting in 1174 objects with host absolute magnitudes, and 1195 objects with M$_*$ estimates. These include SNe~II and SESNe (SNe IIb, Ib, Ic, Ic-BL, Ibc). A smaller number of SNe also have oxygen abundance (metallicity) measurements at their sites of explosion. These abundances were derived by measuring the fluxes of H$\alpha$, H$\beta$, [O~III] $\lambda5007$, and [N~II] $\lambda6583$ and applying the O3N2 and N2 diagnostic calibrations from \citet{Pettini04}. Specifically, 412 SNe have N2 measurements, and 296 SNe have O3N2 values. The number of objects per SN type is summarised in Table~\ref{tab:litsample}.

\begin{figure*}
\centering
\includegraphics[width=\columnwidth]{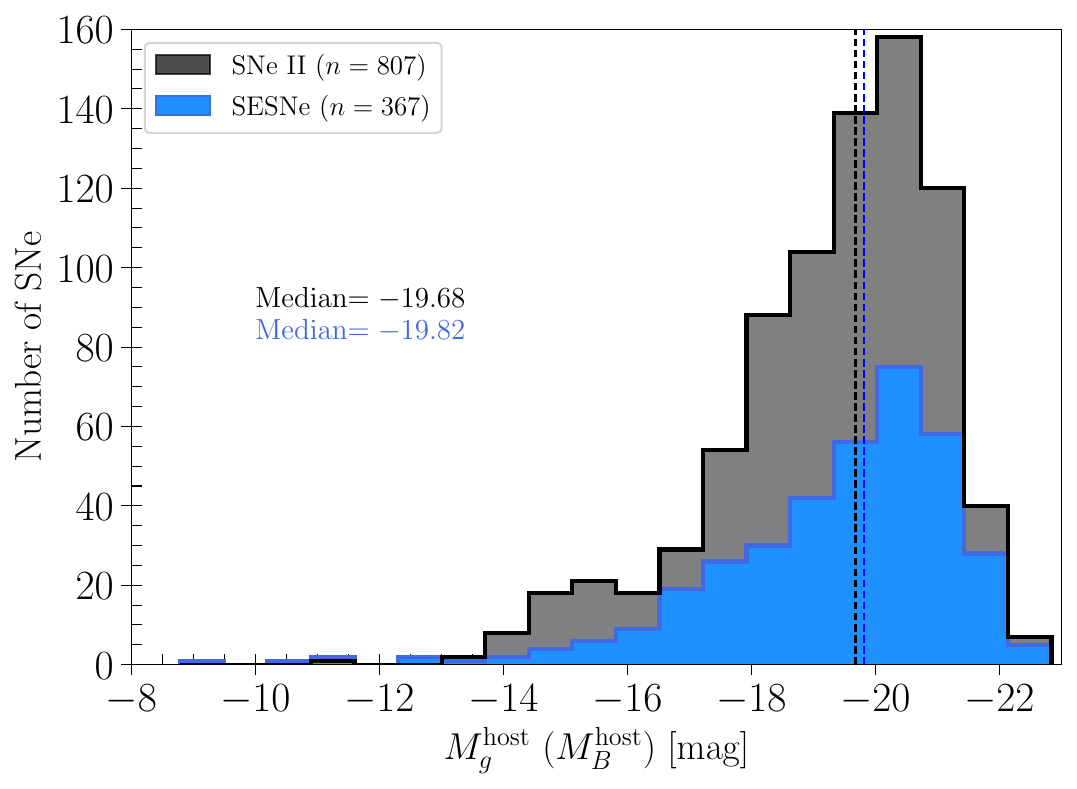}
\includegraphics[width=\columnwidth]{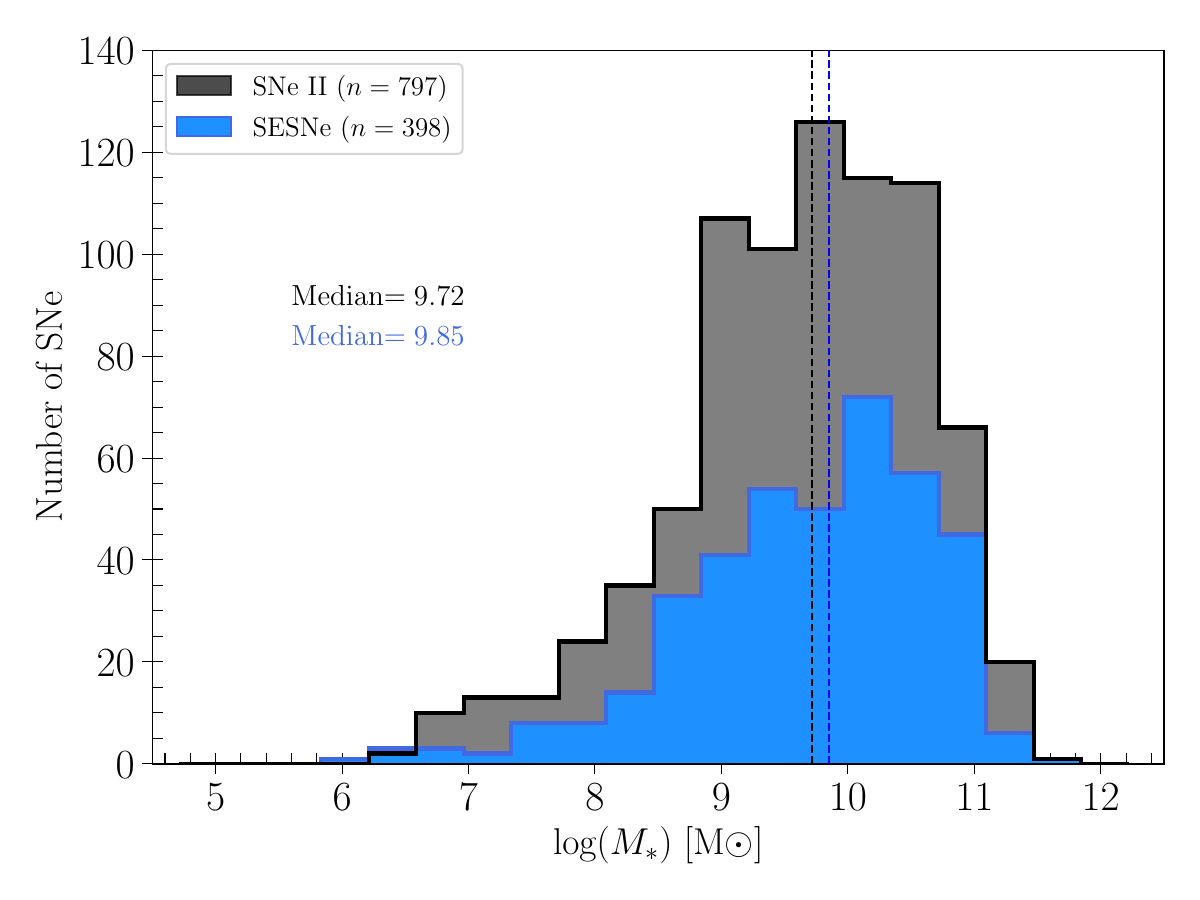}
\caption{Distribution of the host galaxy absolute magnitudes (left) and stellar masses (right) for the literature sample. The vertical lines indicate the median for the samples shown in the legend.}
\label{fig:histograms}   
\end{figure*}

Figure~\ref{fig:histograms} shows the distributions of M$_g^{host}$ (M$_B^{host}$) and M$_*$ for the literature sample. SESNe are predominantly found in more massive ($\overline{log(M_*)}=9.85$ \Msun) and brighter galaxies ($\overline{M_g^{host}}=-19.82$ mag) compared to SNe~II ($\overline{log(M_*)}=9.72$ \Msun\ and $\overline{M_g^{host}}=-19.68$ mag). The sample also shows a clear deficit of SNe in very faint, low-mass galaxies, primarily because most literature data are biased toward SNe discovered in bright, targeted galaxies.

\begin{figure}
\centering
\includegraphics[width=\columnwidth]{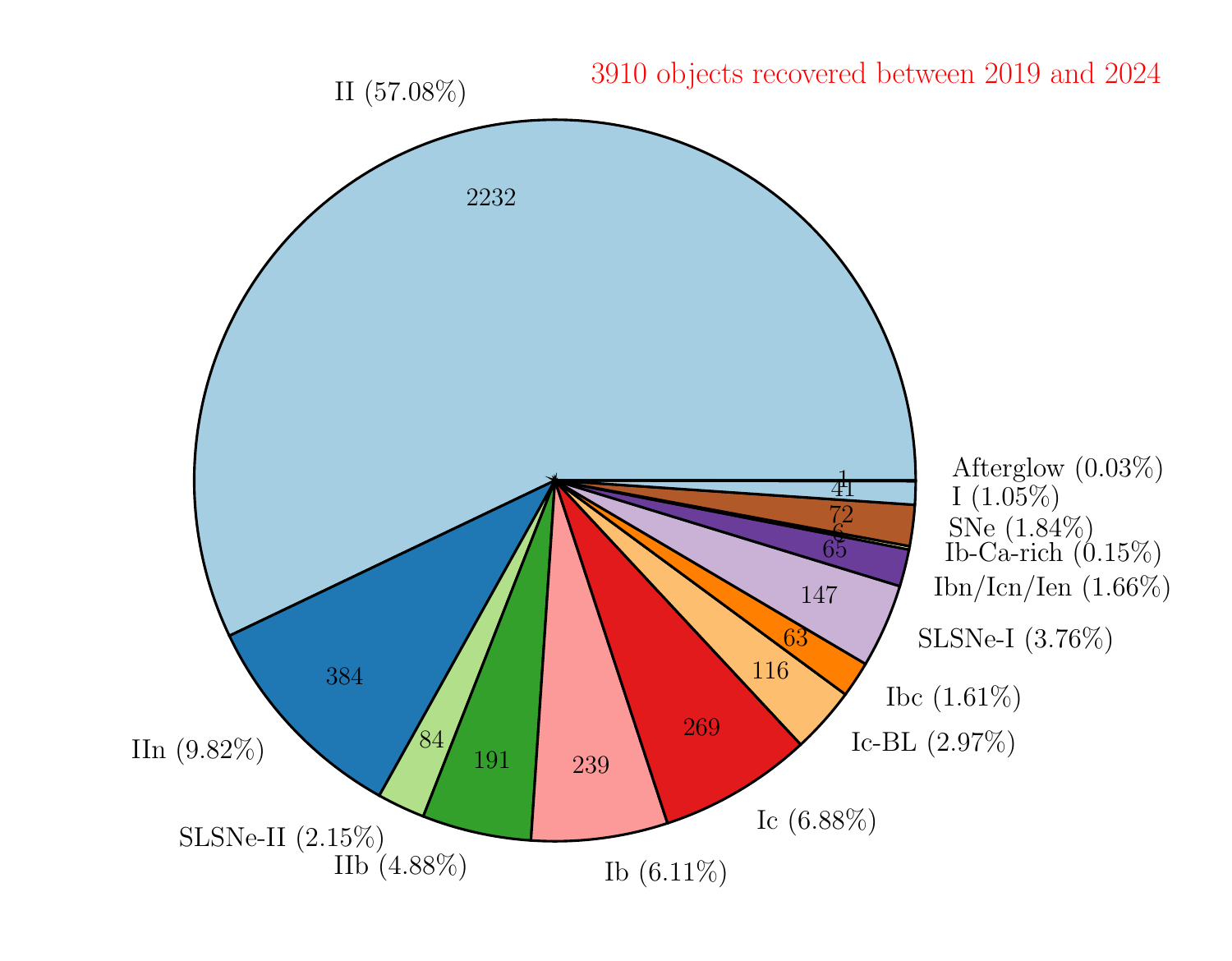}
\vspace{-0.7cm}
\caption{Numbers and relative fractions of SNe (excluding SNe~Ia) classified between 2019 to 2024 and reported to TNS.}
\label{fig:pie}   
\end{figure}

In the second approach, based on SN classified between 2019 and 2024, we mitigate the biases introduced by the literature sample. Since the Zwicky Transient Facility (ZTF; \citealt{Bellm19, Graham19}) began full operations in 2018, nearly all transient discoveries and classifications have been systematically reported to the TNS from approximately 2018–2019 onward, providing a more homogeneous and comprehensive sample. Moreover, most modern transients originate from untargeted, wide-field surveys such as All-Sky Automated Survey for SuperNovae (ASAS-SN; \citealt{Shappee14}); the Asteroid Terrestrial-impact Last Alert System (ATLAS; \citealt{Tonry18, Smith20}); and ZTF, which offer precise astrometry and enable robust SN–host associations. Consequently, this new sample minimises selection, classification, and host-association biases, providing a more uniform, statistically reliable, and representative view of the CCSN population across diverse galactic environments. Following the above, we recovered 3910 objects. The corresponding numbers and relative fractions of SNe (excluding thermonuclear Type Ia SNe) are shown in Figure~\ref{fig:pie}.

\begin{figure}
\centering
\includegraphics[width=\columnwidth]{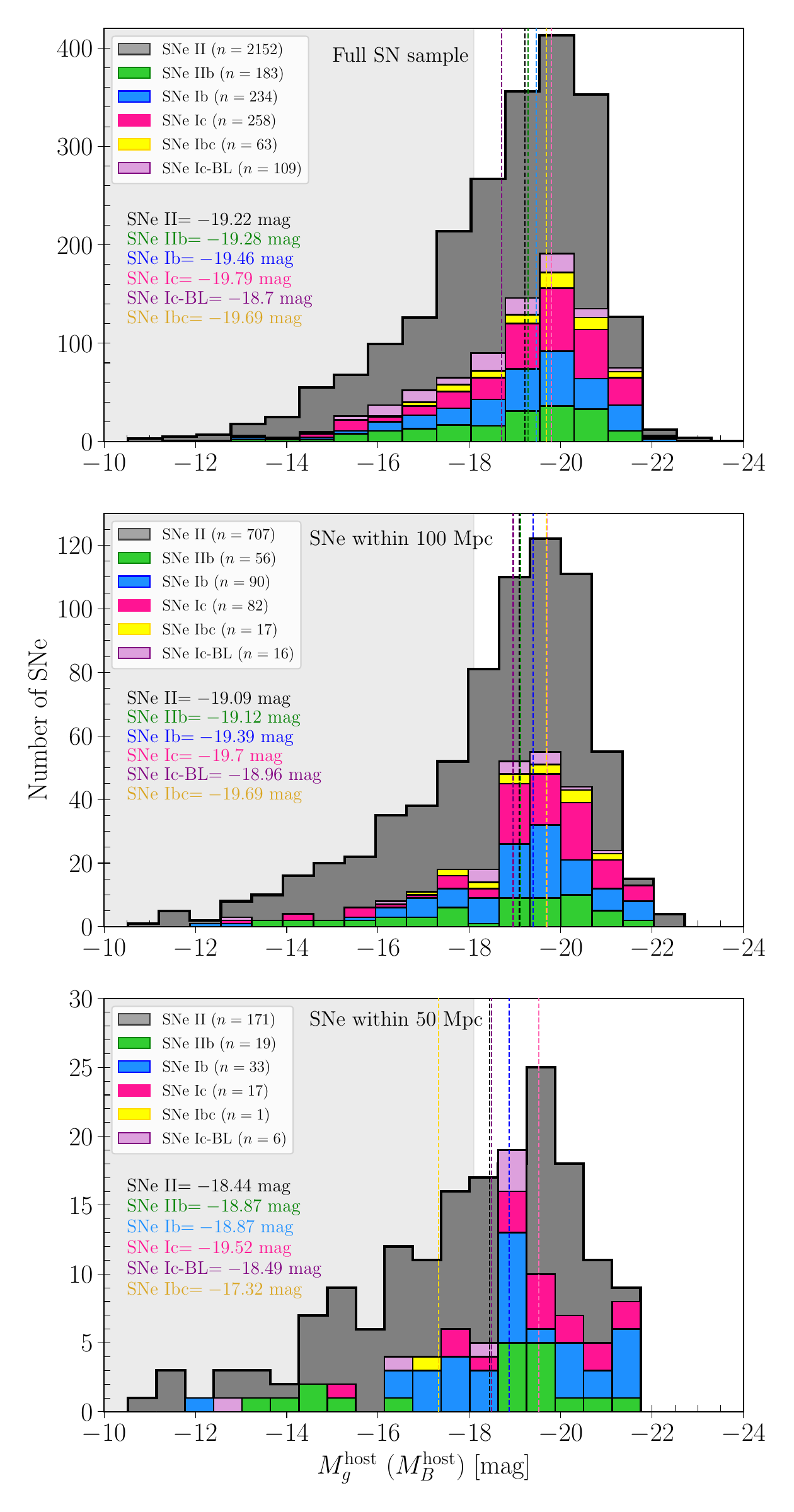}
\caption{Distribution of the host galaxy absolute magnitudes for the sample obtained between 2019 and 2024. \textbf{Top:} Full sample. \textbf{Middle:} SNe within 100 Mpc. \textbf{Bottom:} SNe within 50 Mpc. The vertical lines indicate the median for the samples shown in the legend. The grey shaded area highlights the low-luminosity region (M$_g^{host}$ (or M$_B^{host}$)$\gtrsim-18.5$ mag).}
\label{fig:histoTNS}   
\end{figure}

From this dataset, we select a subset of SN types (SNe~II, IIb, Ib, Ic, Ic-BL, Ibc) for which a host galaxy could be reliably identified and photometry obtained. Host associations were established using the SIMBAD astronomical database \citep{Wenger00}, the ALADIN interactive sky atlas \citep{Bonnarel00}, and visual inspection for faint hosts or hostless objects. Photometric data in the $B$ and/or $g$ bands were collected from the Sloan Digital Sky Survey (SDSS) Data Release 16\footnote{\url{https://skyserver.sdss.org/dr16/en/home.aspx}}, the Panoramic Survey Telescope and Rapid Response System (Pan-STARRS;\footnote{\url{https://panstarrs.stsci.edu/}} \citealt{Flewelling16, Chambers16}) data archive and the HyperLEDA\footnote{\url{http://leda.univ-lyon1.fr/}} \citep{Makarov14} database. All magnitudes were corrected for Milky Way extinction following \citet{Schlafly11} and the \citet{Cardelli89} reddening law with R$_V=3.1$. The number of objects in each SN subtype with available M$_g^{host}$ (M$_B^{host}$) is summarised in Table~\ref{tab:SNTNS} and their respective M$_g^{host}$ (M$_B^{host}$) distributions are shown in Figure~\ref{fig:histoTNS}.

\begin{table}
\small
\centering
\caption{Number of objects with host magnitude measurements obtained from 2019 and 2024.}
\label{tab:SNTNS}
\begin{tabular}{lccccc}
\hline
\hline
SN type   & M$_g^{host}$$^{\star}$ & M$_g^{host}$ (100 Mpc)  & M$_g^{host}$ (50 Mpc)\\
          &    [mag]               &         [mag]           &           [mag]      \\
\hline
SNe~II    &          2152          &          707           &         171           \\
SNe~IIb   &           183          &           56           &          19           \\  
SNe~Ib    &           234          &           90           &          33           \\
SNe~Ic    &           258          &           82           &          17           \\
SNe~Ibc   &            63          &           17           &           1           \\
SNe~Ic-BL &           109          &           16           &           6           \\
\hdashline 
Total     &          2999          &          968           &         247           \\
\hline
\end{tabular}
\begin{list}{}{}
\item \textbf{Notes:} 
\item For the 100 Mpc and 50 Mpc subsamples, the cuts in redshift are z$\leq0.023$ and z$\leq0.012$, respectively.
\item $^{\star}$ For the host absolute magnitude, we use either M$_g^{host}$ or M$_B^{host}$. When both measurements are available, we adopt M$_g^{host}$ as the reference value.
\item - Column 3 and 4 are the number of objects with absolute magnitude measurements within 50 and 100 Mpc, respectively.
\end{list}
\end{table}

Given that this \textit{new} sample is neither volume- nor magnitude-limited (as it combines objects from various surveys), we estimate its relative completeness based on the faintest expected CCSN luminosities. Assuming an absolute magnitude of $-15$ mag (typical of low-luminosity SNe~II; \citealt{Pastorello06, Spiro14, Anderson14}), and a more conservative limit of $-16.5$ mag, we estimate the completeness range of our sample based on the spectroscopic completeness of the ZTF and ATLAS surveys. Although ATLAS is generally considered complete to a magnitude of $\sim19.0$ \citep{Tonry18} and ZTF to about $\sim20.0$ \citep{Bellm19}, both surveys are spectroscopically completed only to brighter limits, approximately $m<18.5$ mag \citep{Perley20}. Under these assumptions, the sample is complete within distances of roughly D$\sim50$ Mpc ($z\sim0.012$) for an absolute magnitude of $-15$ mag, and D$\sim100$ Mpc ($z\sim0.023$), for an absolute magnitude of $-16.5$ mag. A distance cut of 50 Mpc is consistent with volume-limited samples adopted in previous studies (\citealt{Li11a}; also see \citealt{Ma25}). The number of objects satisfying these criteria is reported in Table~\ref{tab:SNTNS}. As seen, our \textit{new} sample consists of 247 SNe within 50 Mpc and 968 SNe within 100 Mpc. Figure~\ref{fig:histoTNS} (middle and bottom panels) shows the M$_g^{host}$ (M$_B^{host}$) distributions for these SNe. As expected, the distribution for nearby objects (within 50 Mpc) shifts toward fainter host galaxies compared to the 100 Mpc sample.

\section{Analysis and results}
\label{sec:analysis}

In this section, we compare (i) metallicity estimates derived using different methods, and (ii) relative ratios with results from previous studies. The metallicity comparison is performed exclusively using literature data, while the relative ratio analysis is divided into three parts: one based on values compiled from the literature, and the other consisting of SNe classified between 2019 and 2024, limited to distances of 50 Mpc and 100 Mpc.

\subsection{Metallicity estimators}
\label{sec:metalcomp}

\begin{figure}
\centering
\includegraphics[width=\columnwidth]{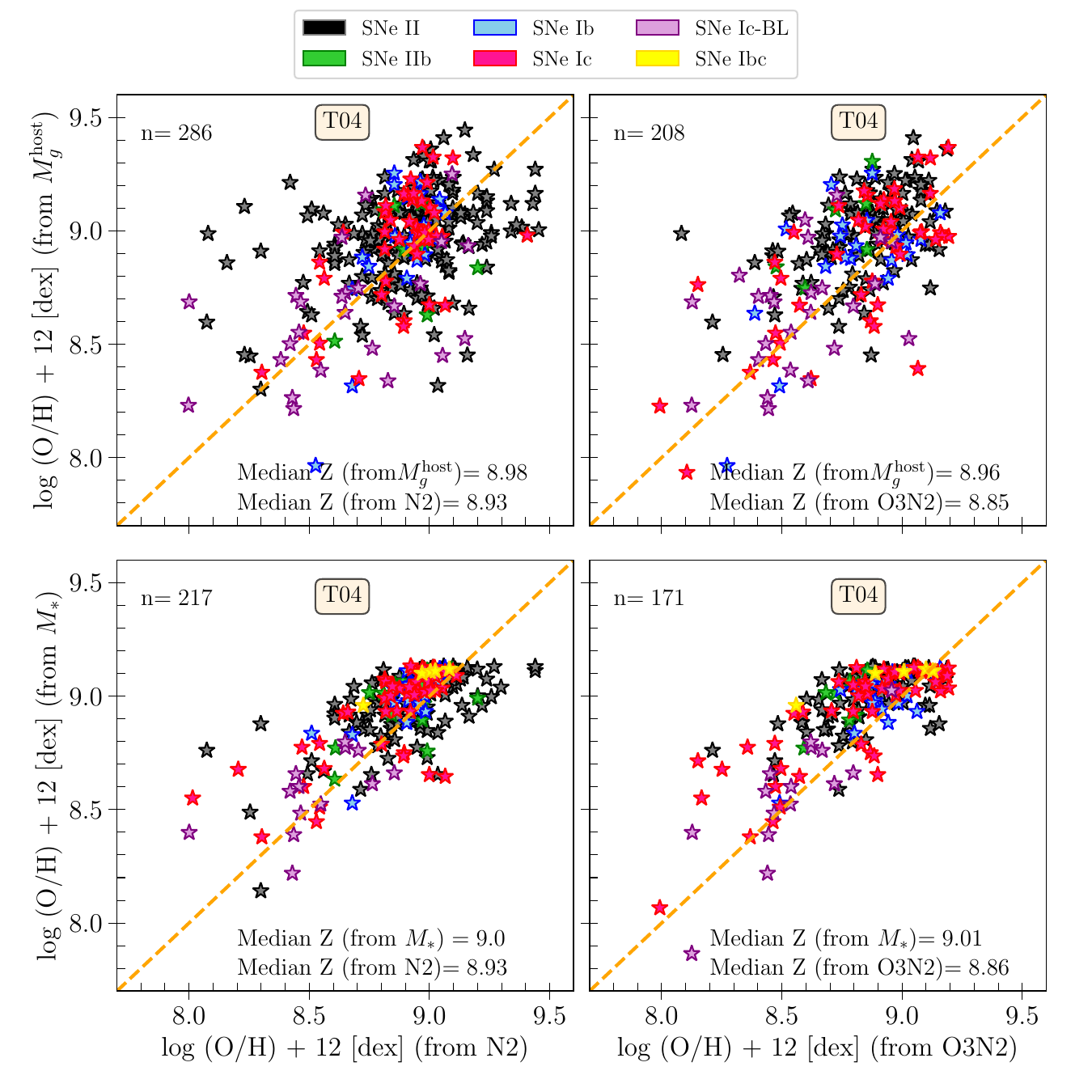}
\caption{Metallicity inferred from M$_g^{host}$ (M$_B^{host}$; top) and M$_*$ (bottom) using the relations of \citet{Tremonti04} versus metallicities derived from PP04 N2 (left) and O3N2 diagnostics (right). The latter have been converted to the \citet{Tremonti04} calibration using the conversion relations of \citet{Kewley08}. The number of events in each panel, along with the median values of the x- and y-axis distributions, are indicated. The orange dashed line denotes the one-to-one ($x=y$) relation. }
\label{fig:metalcomp}   
\end{figure}

Taking advantage of the large number of SNe with multiple host-galaxy parameters available in the literature, we can investigate how metallicity estimates differ when they are derived using various methods. In particular, we compare the oxygen abundances at the SN explosion sites obtained from the O3N2 and N2 diagnostics of \citet[][hereafter PP04 O3N2 and PP04 N2]{Pettini04} with global metallicity estimates inferred from the luminosity–metallicity and mass–metallicity relations of \citet[][hereafter T04]{Tremonti04}. To enable a consistent and meaningful comparison, the PP04 O3N2 and PP04 N2 metallicities are converted to the \citet{Tremonti04} calibration using the conversion relations provided by  \citet{Kewley08}. As shown in Figure~\ref{fig:metalcomp}, metallicities derived from global host parameters are slightly higher than those measured directly at the SN explosion sites. Furthermore, metallicities inferred from M$_g^{host}$ display a larger scatter than those obtained from M$_*$. While estimates based on M$_g^{host}$ can reach values of 12 + log(O/H)$\sim9.5$ dex, metallicities inferred from the mass–metallicity relation exhibit a sharp upper limit around 12 + log(O/H)$\sim9.2$ dex, which, in turn, produces an apparent saturation (see Section~\ref{sec:discussion}). 
 
Across most of the metallicity range, estimates derived from M$_*$ and  M$_g^{host}$ show a larger dispersion at the lower-mass/fainter luminosity end. More precisely, a subset of objects (SNe~II and Ic) exhibits higher global metallicities than the corresponding local measurements at 12 + log(O/H)$<8.5$ dex.

\begin{figure}
\centering
\includegraphics[width=\columnwidth]{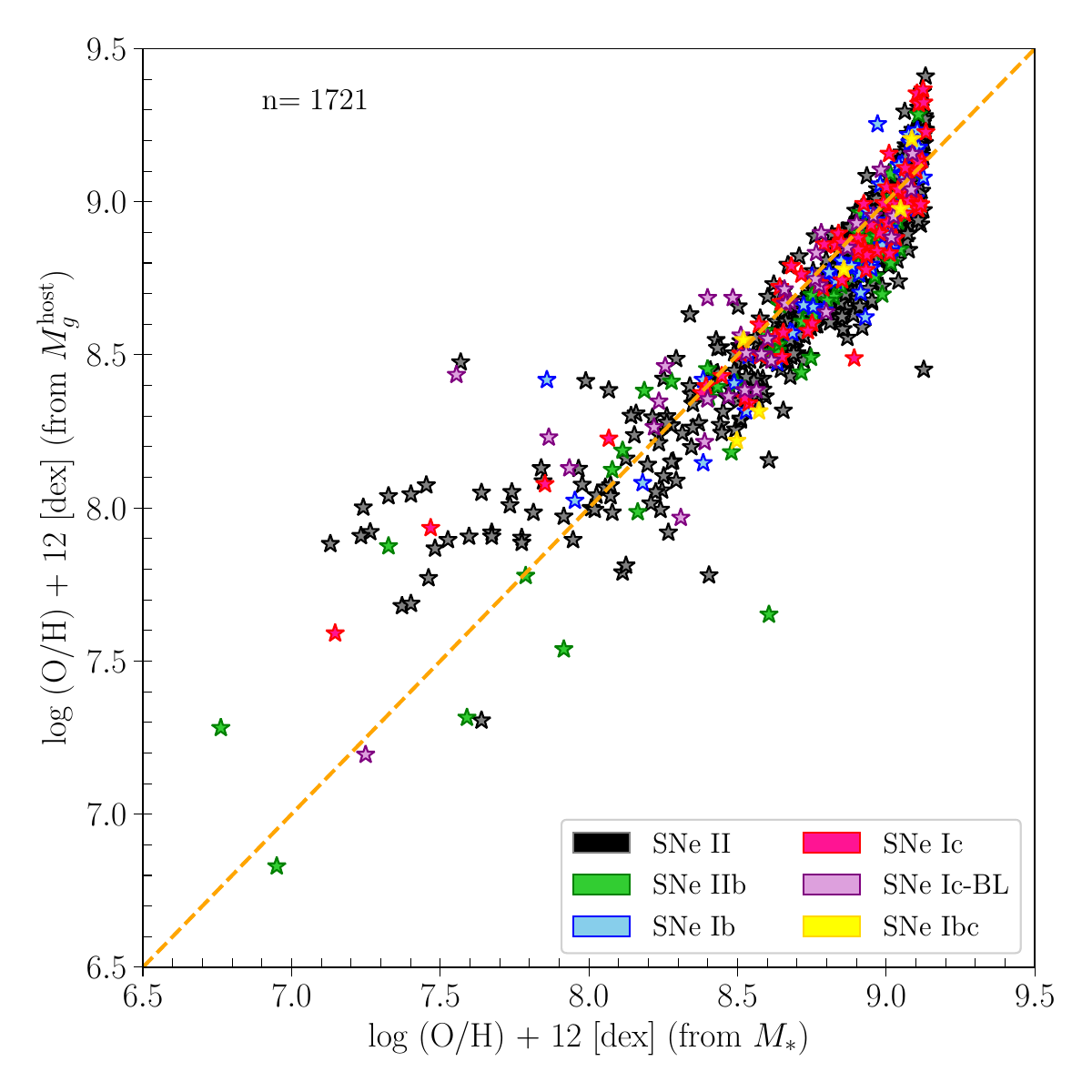}
\caption{Metallicity inferred from M$_g^{host}$ (M$_B^{host}$) and M$_*$ using the relations of \citet{Tremonti04}. The orange dashed line denotes the one-to-one ($x=y$) relation.}
\label{fig:metalcompg}   
\end{figure}

When comparing the metallicity inferred from the luminosity–metallicity and mass–metallicity relations (Figure~\ref{fig:metalcompg}), the sharp upper limit around 12 + log(O/H)$\sim9.2$ dex (set by the M$_*$) is clearly seen (see Section~\ref{sec:discussion}). Despite some scatter, the metallicity derived from M$_*$ and M$_g^{host}$ shows good agreement over the range 8.0$<$12 + log(O/H)$<$9.2 dex. However, for lower-mass galaxies (log(M$_*)<8.0$ \Msun), the values inferred from M$_g^{host}$ (M$_B^{host}$) tend to be systematically higher (by 0.1 -- 0.8 dex) than those inferred from M$_g^{host}$. For instance, at 12 + log(O/H)$\sim7.1$ dex inferred from M$_*$, the corresponding 12 + log(O/H) from M$_g^{host}$ reaches approximately 7.9 dex. In addition, there is a clear lack of events hosted by faint, low-mass galaxies (see Figure~\ref{fig:histograms}), consistent with the targeted nature of earlier SN searches (see Section~\ref{sec:bias}.

Despite the heterogeneity of the sample (arising from the use of different surveys and methods to estimate host-galaxy properties), the comparisons indicate that the resulting metallicity measurements are largely consistent across the majority of objects. This overall agreement indicates that, while some systematic offsets may be present between subtypes (particularly for SNe in low/faint host), these do not substantially impact the global expected trends: metallicity increases systematically with galaxy luminosity and stellar mass.

\subsection{Observational bias or intrinsic deficit?}
\label{sec:bias}

\begin{figure}
\centering
\includegraphics[width=\columnwidth]{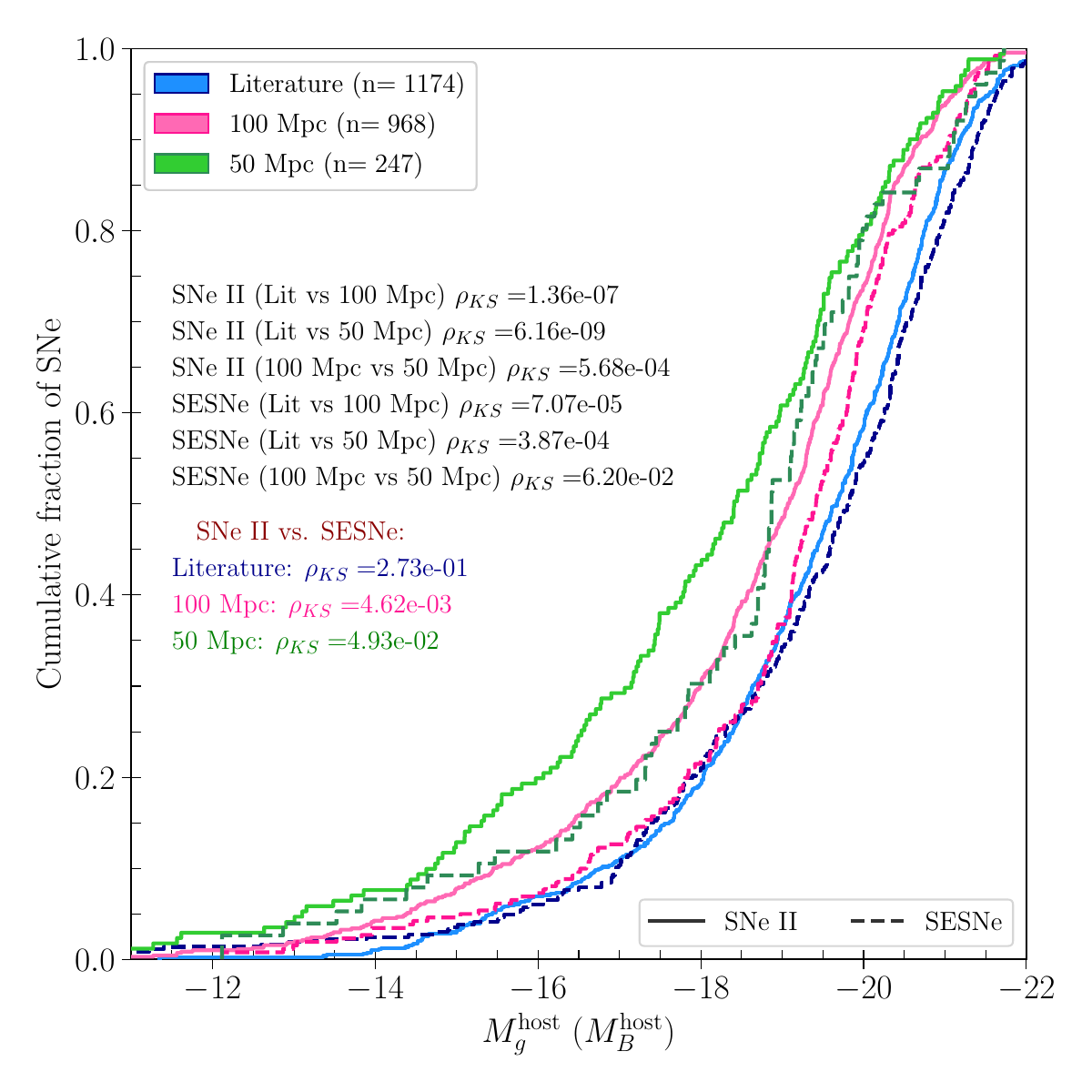}
\caption{Cumulative distributions of host-galaxy absolute magnitudes M$_g^{host}$ (M$_B^{host}$) for SNe~II (continuous line) and SESNe (dashed line) across the literature (blue), 100 Mpc (pink), and 50 Mpc (green) samples. The number of events in each sample is indicated, along with the results of KS tests comparing the different SN populations.}
\label{fig:dist}   
\end{figure}

To assess whether the apparent lack of CCSNe in faint, low-mass galaxies in the literature sample (Figures~\ref{fig:histograms} and \ref{fig:metalcompg}) arises from observational bias or reflects an intrinsic distribution, we compare the host-galaxy absolute-magnitude distributions of the literature, 100 Mpc, and 50 Mpc subsamples. The literature sample is known to be biased toward brighter galaxies, as many of its SNe were discovered in targeted surveys. In contrast, the 100 Mpc and 50 Mpc subsamples are identified primarily from wide-field, untargeted surveys and therefore are less affected by this bias. To quantify the statistical significance of any differences between these samples, we employ the Kolmogorov–Smirnov (KS) test \citep{ks-test1,ks-test2}, which evaluates whether two samples are drawn from the same parent distribution. We reject this null hypothesis when $\rho_{KS}<0.05$. 

Figure~\ref{fig:dist} presents the distribution of host-galaxy absolute magnitudes (M$_g^{host}$ or M$_B^{host}$) for SNe~II and SESNe across three samples: the literature compilation and the 100 Mpc and 50 Mpc subsamples. As shown in the figure, there are significant differences between the literature sample and the untargeted (100 and 50 Mpc) subsamples for both SNe~II and SESNe. In particular, CCSNe from the literature sample are preferentially hosted by more luminous galaxies, while CCSNe from untargeted surveys span a broader range of host luminosities, including a substantial fraction in fainter galaxies.

For SNe~II, the difference between samples is most pronounced when comparing the literature and 50 Mpc subsample ($\rho_{KS}=1.36\times10^{-7}$), while for SESNe the largest difference is found between the literature and the 100 Mpc subsample ($\rho_{KS}=7.07\times10^{-5}$). We also find significant differences between the host galaxy luminosity distributions of SNe~II and SESNe within 100 Mpc ($\rho_{KS}=4.62\times10^{-3}$) and 50 Mpc ($\rho_{KS}=4.93\times10^{-2}$). These results indicate that, on average, SNe~II tend to occur in less luminous host galaxies than SESNe.

\subsection{Relative ratios as a function of metallicity}
\label{sec:ratios}

To investigate the dependence of SN type on metallicity, \citet{Prantzos03} analysed the observed ratio of SNe~Ibc to SNe~II as a function of galaxy absolute magnitude. They found a strong correlation, with SNe Ibc occurring preferentially in higher-metallicity environments (brighter galaxies) compared to SNe~II. Several years later, \citet{Prieto08}, using the N2 and O3N2 diagnostics, confirmed this trend.

Following the methodology presented in \citet{Prieto08}, we calculate these measurements using our dataset. Specifically, we compute the ratio of SESNe-to-SNe~II in bins containing an equal number of SNe, adopting a total of five bins in each case. This binning strategy ensures comparable statistical weight and more uniform uncertainties across bins, while reducing biases arising from the irregular sampling of the underlying distributions. Using the same binning also allows for a direct and consistent comparison with previous studies. Poisson uncertainties are adopted for the SNe ratios, while metallicity uncertainties are represented by the metallicity range covered by each bin. In this analysis, SESNe include the IIb, Ib, and Ic subtypes. In addition to the combined SESNe-to-SNe II ratio, we also compute subtype-specific ratios, namely IIb/II, Ib/II, and Ic/II.

\subsubsection{Literature sample}

\begin{figure*}
\centering
\includegraphics[width=\textwidth]{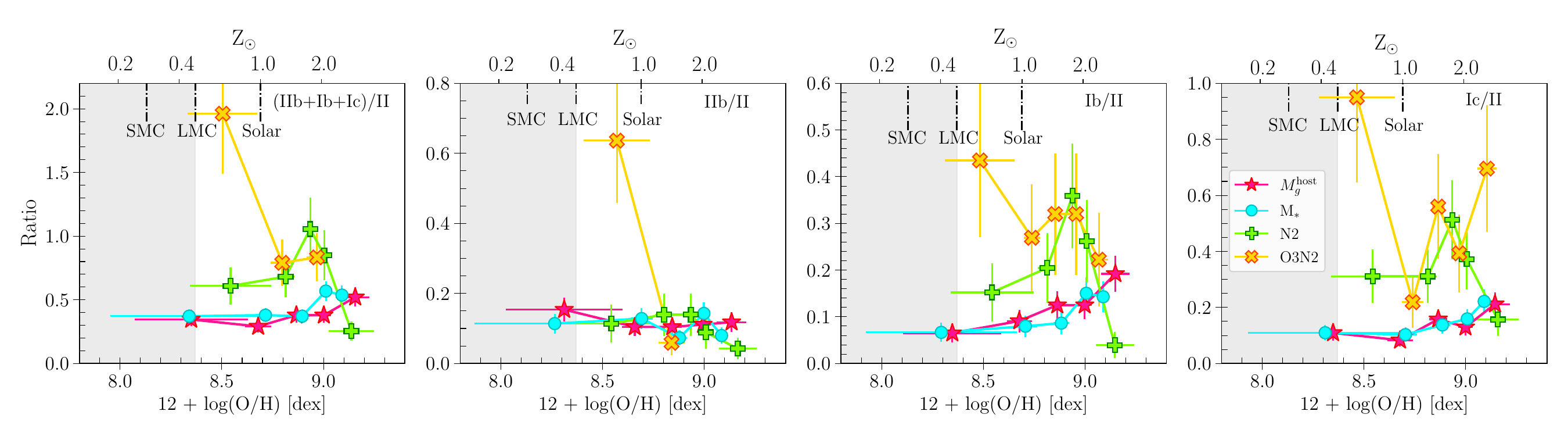}
\caption{Comparison of relative SN type ratios from the literature sample as a function of metallicity, inferred from both global (M$_g^{host}$ as magenta stars, M$_*$, as cyan circles) and local (N2 as green crosses and O3N2, as gold x) galaxy properties. The ratio configuration for each case is indicated in the top right of the corresponding panel. Vertical black dashed lines mark the metallicities of the Small Magellanic Cloud (SMC), the Large Magellanic Cloud (LMC), and the Sun for reference. The grey area highlights the low-metallicity regime (12 + log(O/H)$<8.37$), defined using the metallicity of the LMC.}
\label{fig:ratioscomp}   
\end{figure*}

Figure~\ref{fig:ratioscomp} presents a comparison of these four ratio configurations as a function of metallicity, derived globally from M$_g^{host}$, M$_*$, and locally from N2 and O3N2 diagnostics, and converted to the T04 calibration. Across all configurations, the ratios derived from M$_g^{host}$ and M$_*$ show very similar behaviour. Focusing on the ratios obtained from the M$_g^{host}$, we find that in most cases, the ratios increase with metallicity. The only exception is the IIb/II ratio, which instead shows a slight decrease and then an almost flat trend toward higher metallicities. 

In contrast, the ratios derived using the N2 and O3N2 diagnostics differ from each other. At low metallicity, the O3N2-based ratios are systematically higher, although the corresponding uncertainties in this bin are generally larger than in the others, so these results should be interpreted with caution. On the contrary, the N2-based generally exhibit lower values at higher metallicities. Overall, the trends derived from both diagnostics appear irregular and lack a clear, consistent pattern across the metallicity range.

When comparing ratios derived from local and global metallicities, two main differences appear. First, ratios based on local metallicities are systematically higher than those derived from global values, typically by a factor of $\sim2$. The exceptions are observed in the IIb/II ratio, which remains comparable between all approaches. Second, the metallicity range covered differs notably: ratios from global metallicities extend from $\sim8.3$ dex (with a broader bin) up to $\sim9.2$ dex, while those from local metallicities span a narrower range, from about $\sim8.5$ dex to $\sim9.1$ dex.

\subsubsection{SNe within 50 and 100 Mpc observed between 2019-2024}

\begin{figure*}
\centering
\includegraphics[width=\textwidth]{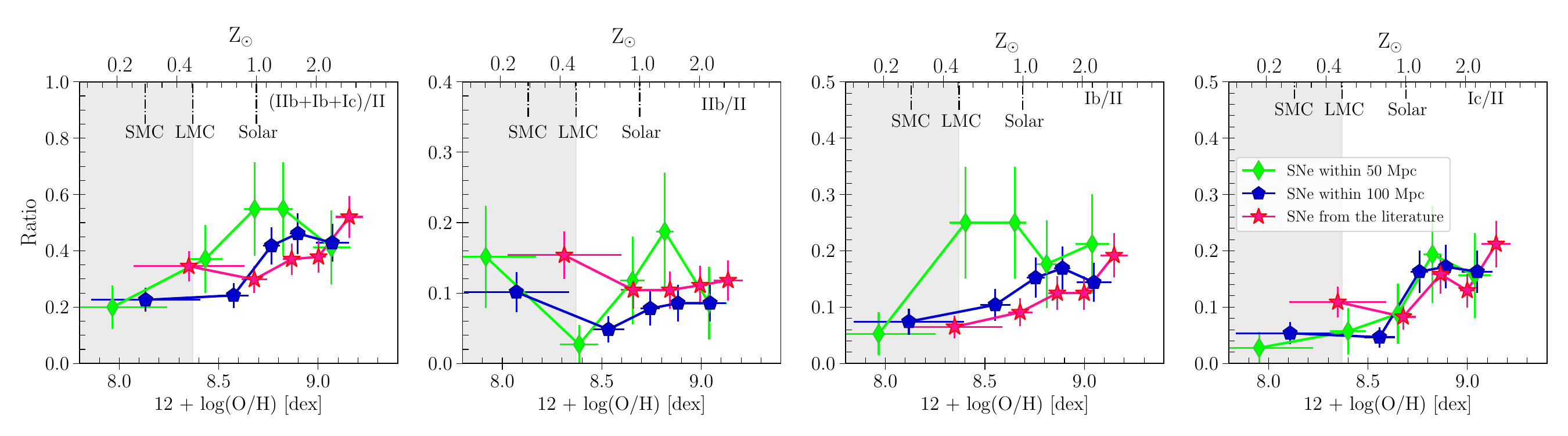}
\caption{Same as Figure~\ref{fig:ratioscomp}, but for SNe within 50 and 100 Mpc classified between 2019 and 2024. Metallicities were inferred from global (M$_g^{host}$) galaxy properties.} 
\label{fig:ratiostnscompfull}   
\end{figure*}

Figures~\ref{fig:ratiostnscompfull} shows a comparison of the four ratio configurations as a function of metallicity for SNe classified between 2019 and 2024, considering two distance-limited subsamples: events within 50 Mpc (green diamonds) and those within 100 Mpc (blue pentagons). For most ratios, the overall trend shows a gradual increase from low values at low metallicities, levelling off around the third or fourth bin, and followed by a flattening or slight decline toward higher metallicities. This behaviour is observed in both samples, though it is more pronounced for the 50 Mpc subset. The main exception is the IIb/II ratio, which exhibits a nearly flat evolution in the 100 Mpc sample and a more irregular pattern among the nearby (50 Mpc) events. Overall, the ratios derived from both samples are broadly consistent with one another.

An important difference emerges at the low-metallicity end: the first bin for the 50 Mpc sample corresponds to lower average metallicities (⟨12 + log(O/H)⟩ $=7.94$ dex) compared to the 100 Mpc sample (⟨12 + log(O/H)⟩$=8.13$ dex). This shift likely reflects the higher fraction of low-luminosity, metal-poor galaxies in the nearby sample, providing a slightly better probe of the metal-poor regime.

\subsubsection{Literature sample vs SNe within 50 and 100 Mpc}

To investigate how a heterogeneous literature sample compares with a more uniform distance-limited dataset (SNe within 50 Mpc and 100 Mpc), we analyse the derived subtype ratios as a function of metallicity. Since for the 50 Mpc and 100 Mpc samples only host galaxy magnitudes (M$_g^{host}$) were measured, and considering that this parameter is also more commonly reported in the literature (i.e. a larger comparison sample), we derive the ratios for all three datasets using metallicities inferred from the luminosity-metallicity relation.

Figures~\ref{fig:ratiostnscompfull} presents the comparison of the four ratio configurations for these three samples: the literature compilation, the 50 Mpc subsample, and the 100 Mpc subsample. Despite the heterogeneous nature of the literature sample, which includes SNe from different sources at varying distances, the overall behaviour of the SN subtype ratios closely resembles that of the distance-limited datasets. However, a few notable differences appear: the metallicity range covered by the literature sample is narrower and biased toward higher values (from 12 + log(O/H)$\sim8.3$ to 9.2 dex), reflecting a lack of events in the low-metallicity (low-luminosity) regime (see Section~\ref{sec:bias}). While the general trend across most ratio configurations is to increase with metallicity, the literature sample displays some irregularities: for instance, in the (IIb+Ib+Ic)/II ratio, the lowest metallicity bin does not correspond to the smallest ratio value; instead, the ratio initially declines and then rises from the second bin onward. In contrast, for the Ib/II and Ic/II ratio configurations, the literature and 100 Mpc samples show good agreement.

\section{Discussion}
\label{sec:discussion}

Using both a literature compilation and a newly constructed sample of SNe within 50 and 100 Mpc, classified between 2019 and 2024, we investigated the relative frequencies of CCSN subtypes as a function of metallicity. In this section, we examine the methods used to derive metallicities and discuss their inherent limitations. We also compare our results with those from previous studies to place them in context and to explore the implications for the diversity and evolutionary pathways of massive star progenitors.

\subsection{Limitations of Metallicity estimates}

As shown in Section~\ref{sec:metalcomp}, metallicities derived from the mass–metallicity relation of \citet{Tremonti04} show a sharp upper limit at 12 + log(O/H)$\sim9.2$ dex that arises from the restricted stellar mass range (8.5$<$log(M$_*$)$<$11.5) adopted in their calibration. As discussed in that study, the relation is roughly linear from 10$^{8.5}$ to 10$^{10.5}$ \Msun, but gradually flattens at higher masses (a trend clearly visible in the lower panels of Figure~\ref{fig:metalcomp}), resulting in a saturation of metallicities estimates at the high-mass end. Although the \citet{Tremonti04} relation also has a lower mass limit ($\sim10^{8.5}$ \Msun), we do not observe a corresponding cutoff in the low-metallicity regime; instead, the metallicity estimates show significant dispersion. Therefore, metallicities derived from M$_*$ in dwarf or very massive galaxies may not be reliable. In contrast, this saturation is not observed when using metallicities derived from the luminosity-metallicity relation (Figures~\ref{fig:metalcomp} and \ref{fig:metalcompg}). As noted in \citet{Tremonti04}, the main distinction between the mass–metallicity and luminosity–metallicity relations lies in the stronger turnover at high metallicity when stellar mass is used as the independent variable.

\begin{figure*}
\centering
\includegraphics[width=0.8\columnwidth]{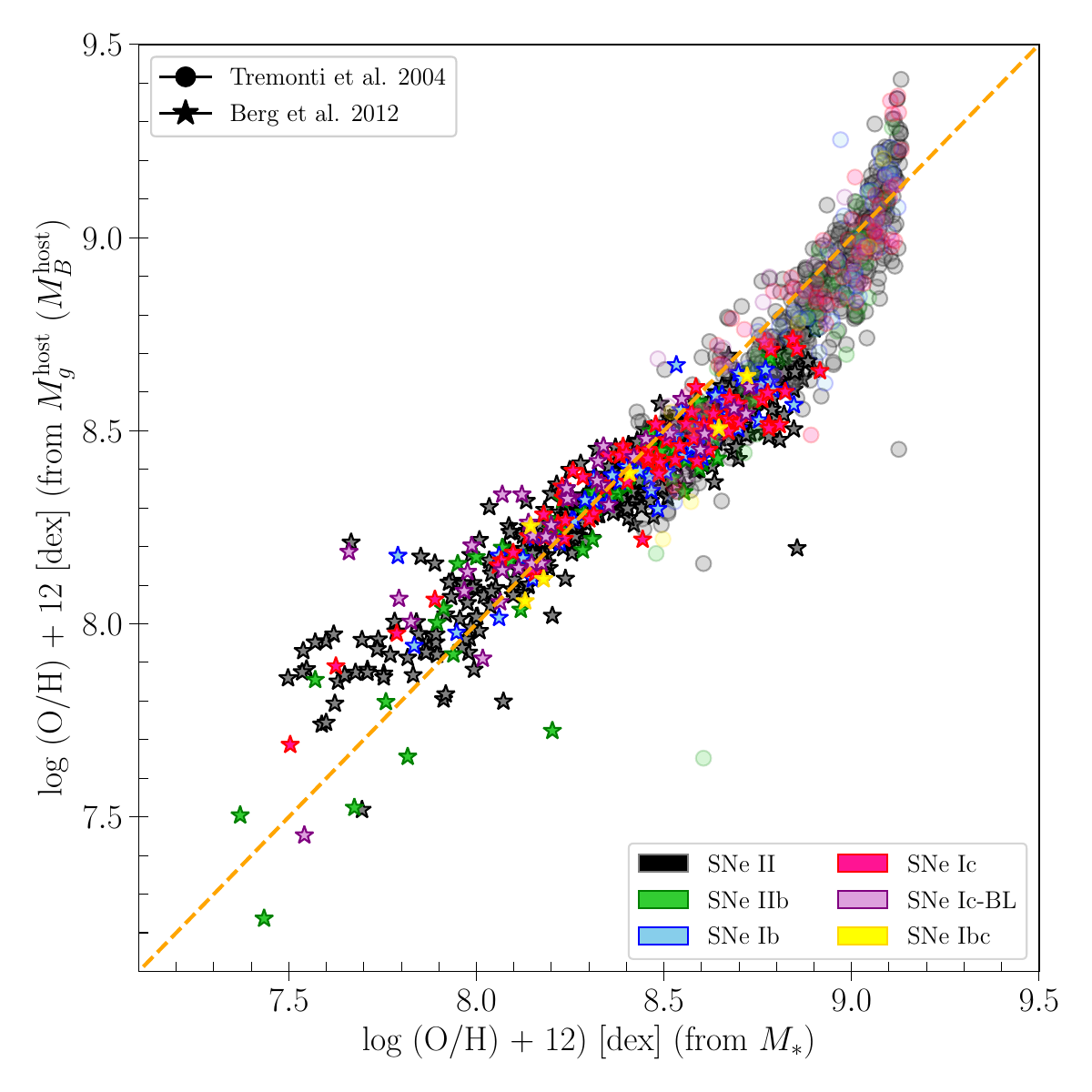}
\includegraphics[width=0.8\columnwidth]{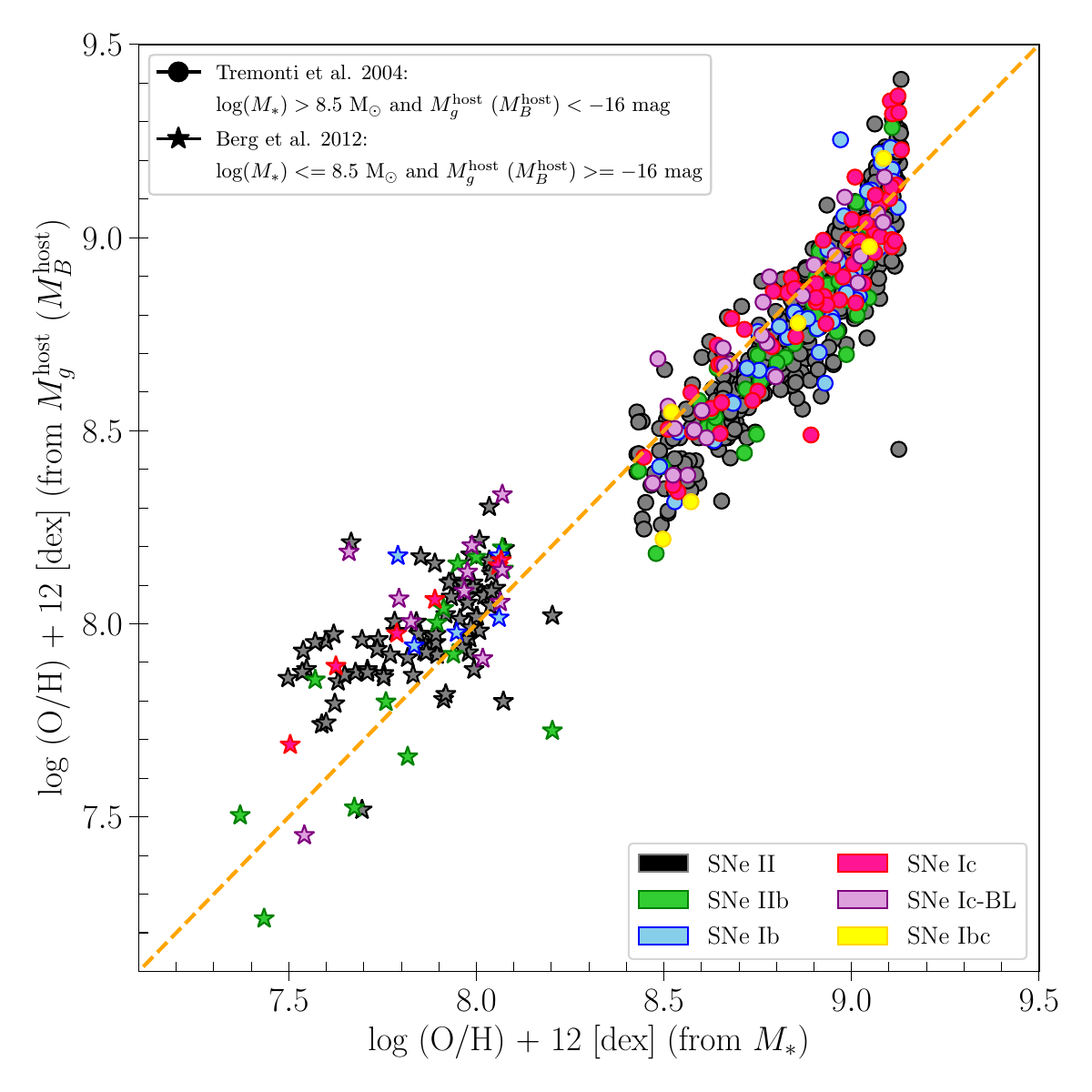}
\caption{Metallicity inferred from M$_g^{host}$ (M$_B^{host}$) and M$_*$ using the relations of \citet[][circles]{Tremonti04} and \citet[][stars]{Berg12}. \textbf{Left:} metallicities estimated with both relations across the full parameter range. \textbf{Right:} metallicities derived within specific ranges of stellar mass and absolute magnitude. The orange dashed line denotes the one-to-one ($x=y$) relation.}
\label{fig:metalTvB}   
\end{figure*}

As is well known, the \citet{Tremonti04} sample is biased toward massive, luminous galaxies. Consequently, a larger dispersion is observed at the low-luminosity end when estimating metallicities from the luminosity–metallicity relation, although the limited number of events prevents a firm conclusion. More recent studies have shown that the slopes of the luminosity–metallicity relation differ for faint galaxies \citep[e.g.][]{Berg12}. This variation in slope among more massive and luminous galaxies may arise from their higher metal and dust content \citep{Rosenberg06}, although its exact origin remains uncertain. The slope differences observed in both the luminosity–metallicity and mass–metallicity relations can significantly impact the derived metallicities, as shown in Figure~\ref{fig:metalTvB}. Metallicities estimated using the \citet{Berg12} relation are systematically lower than those obtained from \citet{Tremonti04}. Selecting the most appropriate calibration is challenging, as each relation lacks complete coverage at either the faint or bright end of the galaxy population. Furthermore, combining both calibrations is not advisable, since the offset between metallicities derived from low-luminosity relations \citep[e.g.][]{Berg12}, and those from \citet{Tremonti04} would introduce an artificial metallicity gap, as shown in Figure~\ref{fig:metalTvB} (right panel). This highlights the need for a more comprehensive sample to refine these relations and provide more reliable constraints on metallicity estimates.

\subsection{Comparison to previous studies: relative ratios for the literature sample}

\begin{figure*}
\centering
\includegraphics[width=\textwidth]{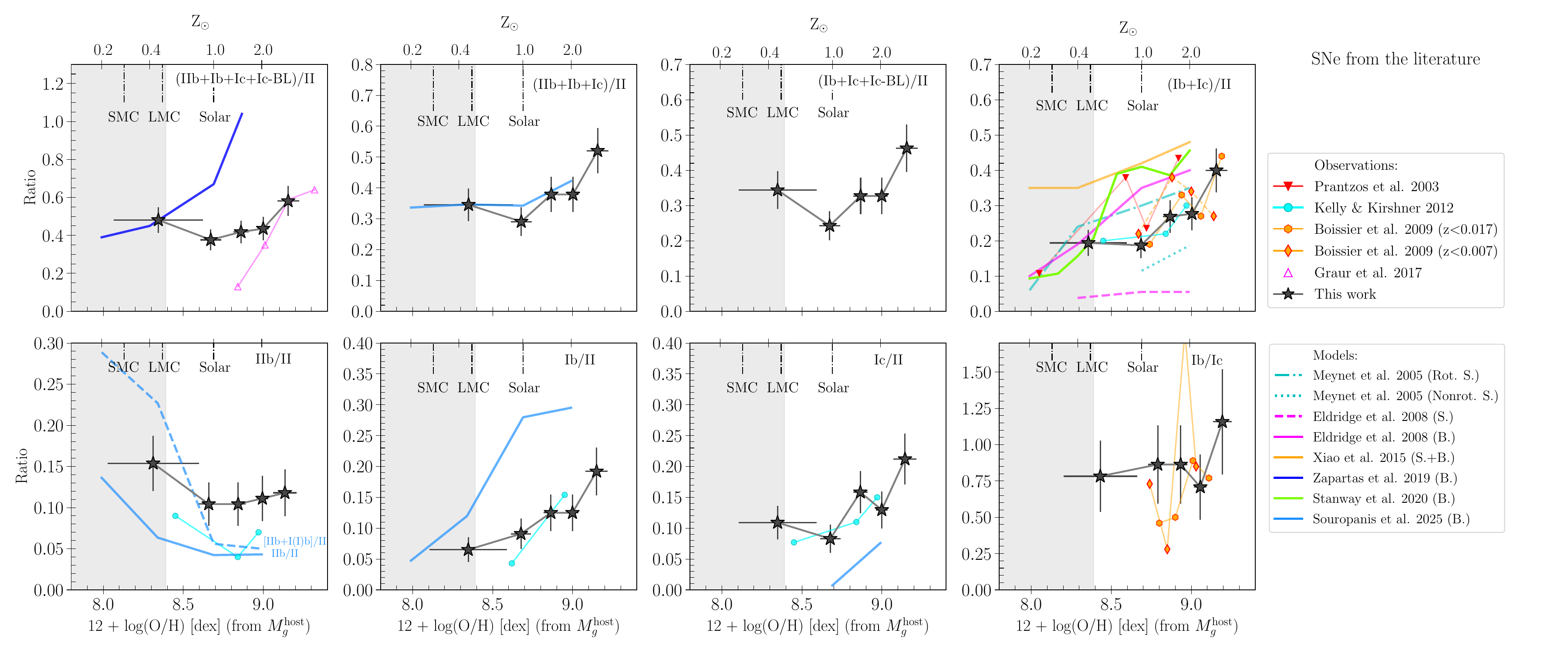}
\caption{Ratio between different SN types as a function of metallicity inferred from M$_g^{host}$ compared with both observations (markers) and models (lines). The ratio configuration for each case is indicated in the top right of the corresponding panel. Vertical black dashed lines mark the metallicities of the SMC, LMC, and the Sun for reference. The grey area highlights the low-metallicity regime (12 + log(O/H)$<8.37$), defined using the metallicity of the LMC.}
\label{fig:ratiosMag}   
\end{figure*}

Acknowledging the limitations of the metallicity estimates, we next examine the relative ratios of different CCSN subtypes as a function of metallicity. Because the number of SNe with local metallicity measurements (N2 and O3N2) is relatively small compared to those with available global estimates, in this section, we focus our analysis on the ratios derived from global measurements, namely M$_g^{host}$\footnote{Given that the ratios derived from M$_*$ behave very similarly to those from M$_g^{host}$ (see Figure~\ref{fig:ratioscomp}), we do not include them here.}. 
Since different studies adopt varying definitions of the SESN class, we calculate the SESNe-to SNe~II ratios under four distinct grouping categories: (IIb+Ib+Ic+Ic-BL)/II; (IIb+Ib+Ic)/II; (Ib+Ic+Ic-BL)/II; (Ib+Ic)/II. Additionally, we consider the individual subtype ratios: IIb/II, Ib/II, Ic/II and Ib/Ic.

Figure~\ref{fig:ratiosMag} shows the resulting relative ratios as a function of metallicity inferred from M$_g^{host}$. To place our results in a broader context, we compare them with previous observational studies, including \citet{Prantzos03, Boissier09, Kelly12} and \citet{Graur17}\footnote{We note that \citet{Graur17} compute the SN ratios by including SNe~IIb in denominator (II+IIb), rather than in the numerator as part of the SESNe).}, which estimated global metallicities using the luminosity-metallicity relation. We also compare our findings with theoretical predictions from both single-star and binary evolution models, including single-star models from \citet[][with and without rotation]{Meynet05}  and \citet{Eldridge08}, binary models from \citet{Eldridge08, Zapartas19, Stanway20} and \citet{Souropanis25}, as well as hybrid single+binary population models from \citet{Xiao15}. Our comparisons do not include the observational results from \citet{Prieto08, Anderson15} and \citet{Kuncarayakti18}, as the metallicities in those studies were derived using the N2 and O3N2 diagnostics for the galaxy centres \citep{Prieto08} and local SN sites \citep{Anderson15, Kuncarayakti18}. 

In the top left panel of Figure~\ref{fig:ratiosMag}, we compare our (IIb+Ib+Ic+Ic-BL)/II ratio with the observational results from \citet{Graur17} and the binary model from \citet{Zapartas19}. Although the model reproduces the general trend of increasing ratios with metallicity, it is offset toward lower metallicity values and predicts systematically higher ratios than our estimates. This discrepancy likely arises from the assumption in \citet{Zapartas19} that all massive stars (M$_{ZAMS}\geq8$ \Msun) explode and produce observable SNe. In contrast, more recent studies \citep[e.g.][]{Souropanis25} allow for the possibility that some massive stars may collapse directly into black holes, producing no or only weak observable transients. As a result, the models of \citet{Zapartas19} predict significantly higher ratios, particularly at higher metallicities. The comparison with \citet{Graur17} shows that their metallicity range is narrower and shifted toward higher values, spanning from 12 + log(O/H)$\sim8.8$ to 9.3 dex, while our sample extends down to $8.3$ dex. Their ratio also increases with metallicity, but with a steeper slope and lower initial values compared to ours. This behaviour may be related to the fact that their ratios do not include SNe IIb in the SESN sample, but instead they are part of the SN~II population.

In the second configuration, (IIb+Ib+Ic)/II, the model from \citet{Souropanis25} reproduces our observed trend well up to its upper limit (2 \Zsun, corresponding to 12 + log(O/H)$\sim9.0$ dex). 
For the (Ib+Ic)/II ratio (the most commonly adopted configuration in the literature; top right panel), the comparison is extensive. Overall, most studies report a similar increasing trend with metallicity. The ratios from \citet{Boissier09} and \citet{Kelly12} agree well with our results, although both cover narrower metallicity ranges: \citet[][for z$<0.017$]{Boissier09}  covers slightly higher metallicity values (12 + log(O/H)$\sim8.7$ to 9.2 dex), while \citet{Kelly12} span a lower range (12 + log(O/H)$\sim8.4$ to 9.0 dex). The ratios from \citet{Prantzos03}  show lower values at low metallicities but rise more steeply toward the high-metallicity end compared to our results.

When comparing with theoretical models, the single-star models from \citet[without rotation]{Meynet05} and \citet{Eldridge08} tend to underestimate the ratio. In contrast, the rotating single-star model from \citet{Meynet05}, the binary models from \citep{Eldridge08} and \citet{Stanway20}, and the hybrid single+binary model of \citet{Xiao15} all overpredict the observed ratios.

To identify which model best reproduces the (Ib+Ic)/II ratio from the literature sample, we use two statistical metrics: the reduced chi-squared \citep[$\chi^{2}$;][]{Chi2} and the Akaike Information Criterion \citep[AIC;][]{AIC}. The reduced $\chi^{2}$ evaluates the goodness of fit by quantifying the deviations between observed and predicted ratios, weighted by the observational uncertainties, with lower values indicating better agreement. The AIC provides a relative measure of model performance by accounting for both goodness of fit and model complexity, penalising overfitting. In this framework, the preferred model minimises both $\chi^{2}$ and AIC. The results are summarised in Table~\ref{tab:tests}, which shows that the single-star model with rotation from \citet{Meynet05} and the binary model from \citet{Eldridge08} provide the best agreement with the observed (Ib+Ic)/II ratio.

\begin{table}
\small
\centering
\caption{$\chi^{2}$ and AIC values obtained from the comparison between theoretical model predictions and the observed (Ib+Ic)/II ratios across the three samples: literature, 100 Mpc, and 50 Mpc.}
\label{tab:tests}
\begin{tabular}{lcccccc}
\hline
\hline
Model    & \multicolumn{2}{c}{Literature} & \multicolumn{2}{c}{100 Mpc} & \multicolumn{2}{c}{50 Mpc} \\
\hline
         &    $\chi^{2}$   &    AIC       &  $\chi^{2}$   &  AIC       &   $\chi^{2}$  &  AIC        \\
\hline
\hline
M05R$^{\star}$ &  3.29     &   $-5.42$    &    1.48      &  $-13.73$   &    0.29       &   $-12.32$   \\
M05NR    &     7.91        &   17.66      &    11.93     &   38.51     &    4.70       &     9.72     \\
E08S     &     21.02       &   83.21      &    19.45     &   76.09     &    5.29       &     12.69    \\
E08B$^{\star}$ &  6.70     &   11.63      &    2.37      &  $-9.30$    &    0.34       &   $-12.07$   \\
X15H     &     20.19       &   79.08      &    17.48     &   66.27     &    8.47       &     28.58    \\
S20B     &     13.26       &   44.44      &    8.21      &   19.91     &    0.49       &    $-11.31$  \\
\hline
\end{tabular}
\begin{list}{}{}
\item \textbf{Notes:} 
\item Models: \textbf{M05R:} single-star model with rotation from \citet{Meynet05}; \textbf{M05NR:} single-star model without rotation from \citet{Meynet05}; E08S: single-star model from \citet{Eldridge08}; E08B: binary model from \citet{Eldridge08}; X15H: hybrid single+binary population models from \citet{Xiao15}; S20B: binary models from \citet{Stanway20}.
\item $^{\star}$ Models that show good agreement with the observed (Ib+Ic)/II ratios across all three samples.
\end{list}
\end{table}

Turning to the individual subtype ratios, both observations and the \citet{Souropanis25} models exhibit a decreasing trend with increasing metallicity. However, both the models from \citet{Souropanis25} and the observations from \citet{Kelly12} show lower IIb/II ratios than those derived in this work. In the model, this outcome may be attributed to the stronger stellar winds assumed in their models, which suppress the formation of a significant number of SNe~IIb at high metallicity. More recent studies, such as \citet{Vink17}, suggest that low-mass helium stars may experience substantially weaker winds. \citet{Gilkis19} further show that, if these reduced wind strengths are adopted once the surface hydrogen fraction drops below $Xs=0.4$, the hydrogen-rich envelope may not be completely stripped. This behaviour contrasts with the higher mass-loss rates prescribed by \citet{Nugis00}, which were employed in the \citet{Souropanis25} models.

We also compared our IIb/II ratios with the [IIb+I(I)b]/II ratios reported by \citet{Souropanis25}\footnote{The I(I)b is not a formal SN class but refers to progenitors with hydrogen ejecta masses of 0.001–0.033 \Msun\ in \citet{Souropanis25} models. The minimum hydrogen mass needed for a Type IIb rather than a Type Ib is uncertain within this range \citep[][]{Hachinger12, Dessart11}, so the final classification depends on the adopted threshold. \citep[see][for more details]{Souropanis25}.}. As shown, the model predicts significantly higher ratios at low metallicities and systematically lower values at high metallicities compared to the observations. For the Ib/II ratio, the \citet{Souropanis25} model predicts significantly higher values than our measurements, while the measurements from \citet{Kelly12} are in good agreement with ours but span a narrower metallicity range. Although the actual values of \citet{Souropanis25} show a difference from our measurements, the observed trends with metallicity are well reproduced (with less frequent SNe~Ib and more common SNe~IIb at low metallicity). In the case of the Ic/II ratio, the increasing trend is consistent with that reported by \citet{Kelly12} and with the model of \citet{Souropanis25}, although the latter predicts systematically lower values. 

Finally, when comparing the Ib/Ic ratio with \citet{Boissier09}, we find noticeably different behaviours. The ratios reported by \citet{Boissier09} for the z$<0.007$ sample are systematically higher and appear somewhat irregular. In contrast, for the z$<0.017$ subsample, the Ib/Ic ratio generally shows an increasing tendency with metallicity, which is consistent with our measurements, although with systematically lower values.

\subsection{Comparison to previous studies: relative ratios for SNe within 50 and 100 Mpc}

\begin{figure*}
\centering
\includegraphics[width=\textwidth]{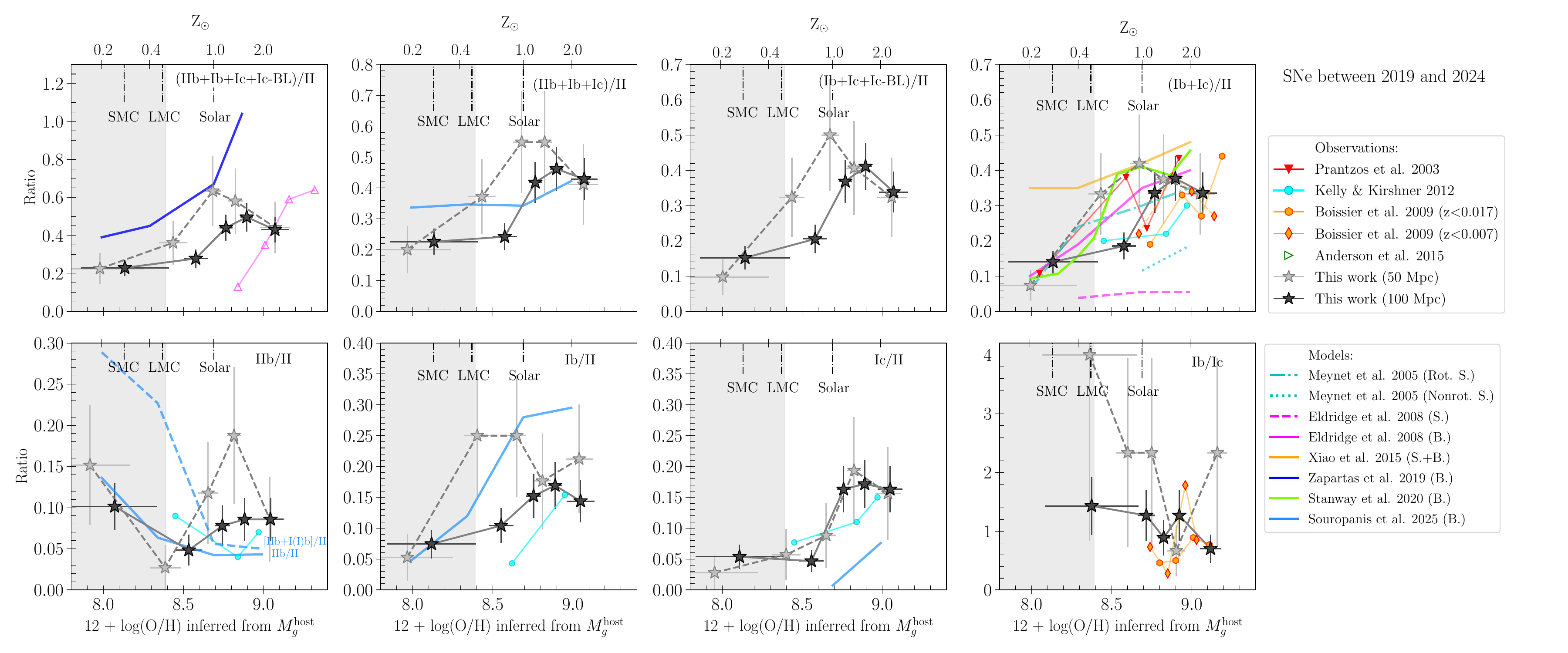}
\caption{Same as Figure~\ref{fig:ratiosMag}, but for SNe within 50 (silver stars) and 100 Mpc (black stars) observed between 2019 and 2024.}
\label{fig:ratiosnew}   
\end{figure*}

In Figure~\ref{fig:ratiosnew}, we present the relative ratios for the 50 Mpc and 100 Mpc subsamples, compared with previous observational studies and theoretical predictions. Since the overall behaviour of different ratios is consistent across the various datasets (literature, 50 Mpc and 100 Mpc), most of the comparisons described in the previous section for the literature sample apply here. Nevertheless, in this section, we focus on the specific features and trends that may differ from those previously discussed.

As observed for the literature sample, the (IIb+Ib+Ic+Ic-BL)/II ratio generally increases with metallicity. This trend is also observed in the \citet{Zapartas19} model and the observations reported by  \citet{Graur17}. However, neither of these fully agrees with our estimates. For the (IIb+Ib+Ic)/II ratio, unlike the literature sample, both 50 Mpc and 100 Mpc subsamples deviate from the \citet{Souropanis25} model. At low metallicities, both the 50 and 100 Mpc subsamples show lower ratios than predicted by the model, but similar values at high metallicities, with the measurements consistent within the error bars.

The (Ib+Ic)/II ratio increases with metallicity in the observational data, in general agreement with most theoretical predictions. A notable exception is the single-star model of \citet{Eldridge08}, which shows an approximately flat trend. Using both $\chi^{2}$ and the AIC statistics to quantify the agreement between models and observations, we find that, the rotating single-star model of \citet{Meynet05} and the binary evolution model of \citet{Eldridge08} provide the best match to the data in both the 100 Mpc and 50 Mpc subsamples, with a better overall agreement for the 50 Mpc sample (see Table~\ref{tab:tests}). In contrast, the non-rotating single-star model of \citet{Meynet05}, the single-star model of \citet{Eldridge08} and the hybrid single+binary model of \citet{Xiao15} show significantly poorer agreement, as indicated by their higher $\chi^{2}$ and the AIC values (Table~\ref{tab:tests}).

For individual subtype ratios, the 50 and 100 Mpc samples show better agreements with the \citet{Souropanis25} model compared to the literature sample. For the IIb/II ratio, the model reproduces our estimates reasonably well at low metallicities for both subsamples; however, from 12 + log(O/H)$\sim8.3$ dex onward, our ratios increasingly exceed the model predictions. For the Ib/II ratio, the \citet{Souropanis25} model agrees relatively well with the 50 Mpc sample up to 12 + log(O/H)$\sim8.8$ dex. In contrast, for the Ic/II ratio, our estimates do not match the model predictions for either subsample. On the other hand, across all individual subtype ratios, the observational trend from \citet{Kelly12} broadly agrees with our 100 Mpc and 50 Mpc measurements, although the ratio value can differ (e.g. Ib/II ratio).

In summary, we find that the SESNe-to-SNe~II ratios derived from the literature, 50 Mpc and 100 Mpc subsamples, show an increase as a function of metallicity. When considering individual subtypes, both Ib/II and Ic/II ratios show a positive correlation with metallicity, while the IIb/II ratio shows a mild decrease with increasing metallicity (see Figure~\ref{fig:ratiostnscompfull}).

\subsection{Ratios from local and global metallicities}

In the previous section, we have focused the analysis on metallicities derived from global galaxy properties. However, if one examines the relative frequencies of CCSN subtype ratios based on local metallicities (N2 and O3N2 diagnostics), they are typically about twice as large as those derived from global measurements (see Figure~\ref{fig:ratioscomp}). This increase in ratio values could reflect either a higher number of SESNe or a lower number of SNe~II. To investigate this, we compare the fractional contributions of SESNe in each sample. In the literature sample with N2 values, 44\% of events are SESNe (i.e. 56\% are SNe~II), while in the O3N2 sample, the SESNe fraction increases to 56\%, indicating roughly proportional differences between the two SN subtypes. In contrast, the sample based on absolute magnitudes shows only 31\% SESNe, about one-third of the total. Thus, the higher ratios observed in local measurements arise primarily from the larger number of SESNe in those samples. This likely reflects a selection bias, as many studies have specifically targeted the local environments of SESNe to better understand their progenitors. However, this over-representation does not reflect the true fraction of SE events. Observationally,  volume-limited \citep{Li11b, Shivvers17} and magnitude-limited \citep{Perley20} samples show that SNe~II constitute roughly $\sim70\%$ of all CCSNe — consistent with our literature and distance-limited samples (50 and 100 Mpc), where SNe II account for 70\%, 71\% and 74\%, respectively. Therefore, although local measurements more accurately trace the immediate chemical environment of the SN progenitor, the available samples remain biased, limiting the robustness of conclusions about SN progenitors. 

When comparing our results with theoretical predictions, we find that only binary evolution models are able to reproduce some of the observed relative frequencies of CCSN subtypes. In particular, the \citet{Souropanis25} model matches several of the observed CCSN fraction configurations reasonably well, while the rotating single-star model of \citet{Meynet05} and the binary evolution model of \citet{Eldridge08} provide the best agreement for the (Ib+Ic)/II ratio. None of the single-star models succeeds in reproducing the observed ratios, suggesting that binary interaction and/or rotation may play a more dominant role than metallicity alone in determining the resulting SN type.

Overall, despite the heterogeneity and selection biases present in current datasets, the relative frequencies of CCSN subtypes remain broadly consistent with less biased samples, lending confidence to the observed metallicity–ratio trends. These results underscore the importance of incorporating binary-evolution channels into models of massive-star evolution to better reproduce the observed diversity of CCSNe.

\section{Conclusions}
\label{sec:conclusions}

In this work, we have presented an analysis of the relative frequencies of CCSNe as a function of metallicity using two complementary samples: (i) literature SNe with host galaxy parameters (absolute magnitudes, stellar masses, and/or oxygen abundances); and (ii) SNe classified between 2019 and 2024 with host magnitude information, including distance-limited subsamples within 50 Mpc and 100 Mpc. Our main conclusions are as follows:

\begin{itemize}

\item CCSNe from the literature are preferentially found in luminous hosts, reflecting the bias of targeted surveys. In contrast, distance-limited subsamples (100 and 50 Mpc) reveal that SNe~II systematically occur in less luminous galaxies than SESNe, possibly indicating a genuine difference in their host environments.

\item A pronounced upper metallicity limit is observed when using values derived from the mass–metallicity relation, consistent with the strong turnover at high mass noted by \citet{Tremonti04}. Therefore, less precise metallicity estimations could be derived in this regime.

\item Most SESNe-to-SNe ratios increase with metallicity, confirming that metal-rich environments host a higher fraction of SESNe.

\item SESNe-to-SNe ratios derived from local metallicities are typically a factor of $\sim2$ higher than those from global estimates. This discrepancy mainly arises from an over-representation of SESNe in samples with local measurements, likely due to targeted studies of SE progenitor environments. These results indicate that SESN-to-SN ratios are strongly dependent on sample selection.

\item Despite sample heterogeneity, the relative frequency trends of CCSN subtypes as a function of metallicity are broadly consistent across the literature compilation and both distance-limited subsamples. The 50 Mpc sample probes slightly lower metallicities, reflecting its inclusion of more low-luminosity, metal-poor galaxies.

\item While metallicity clearly influences the relative occurrence of CCSN subtypes, comparisons with theoretical models indicate that binary interactions play a crucial role in shaping the observed diversity of CCSN subtypes.

\end{itemize}

Overall, these results provide further evidence for a metallicity dependence of SN subtype ratios. At the same time, they highlight the need for larger, more homogeneous SN samples with well-characterised local environments, which will be essential to explore the relative roles of metallicity and binarity in massive star evolution. While metallicity trends are apparent, fully disentangling the effects of binarity would require independent constraints on the binary fraction of progenitor populations, which are currently not available.

\begin{acknowledgements}

We thank the anonymous referee for the comments and suggestions that have helped us to improve the paper.
C.P.G. acknowledges financial support from the Secretary of Universities and Research (Government of Catalonia) and by the Horizon 2020 Research and Innovation Programme of the European Union under the Marie Sk\l{}odowska-Curie and the Beatriu de Pin\'os 2021 BP 00168 programme. 
C.P.G. and L.G. recognise the support from the Spanish Ministerio de Ciencia e Innovaci\'on (MCIN) and the Agencia Estatal de Investigaci\'on (AEI) 10.13039/501100011033 under the PID2023-151307NB-I00 SNNEXT project, from Centro Superior de Investigaciones Cient\'ificas (CSIC) under the PIE project 20215AT016 and the program Unidad de Excelencia Mar\'ia de Maeztu CEX2020-001058-M, and from the Departament de Recerca i Universitats de la Generalitat de Catalunya through the 2021-SGR-01270 grant.

\end{acknowledgements}

\bibliographystyle{aa}
\bibliography{Bibliography}




\end{document}